\documentclass[aps,prc,preprint,showpacs,superscriptaddress,floatfix,raggedbottom]{revtex4-1}
\usepackage{amsmath, amsthm, amssymb}
\usepackage{graphics}
\usepackage{graphicx}
\usepackage{epsfig}
\usepackage{gensymb}        
\usepackage{float}
\usepackage{multirow}   
\usepackage{subfig}    
\usepackage{array}
\begin{document}
\title{Elastic and inelastic scattering of $^{16}$O on $^{27}$Al and $^{28}$Si at 240 MeV}

\author{L.M.~Fonseca}
\author{R.~Linares}
\affiliation{Instituto de F\'isica, Universidade Federal Fluminense, 24210-340, Niter\'oi, Rio de Janeiro, Brazil}
\author{V.A.B.~Zagatto}
\affiliation{Instituto de F\'isica, Universidade Federal Fluminense, 24210-340, Niter\'oi, Rio de Janeiro, Brazil}
\affiliation{Instituto de F\'isica, Universidade de S\~ao Paulo, S\~ao Paulo, Brazil}
\author{F.~Cappuzzello}
\affiliation{Istituto Nazionale di Fisica Nucleare, Laboratori Nazionali del Sud, I-95125 Catania, Italy}
\affiliation{Dipartimento di Fisica e Astronomia "Ettore Majorana", Universit\`a di Catania, I-95125 Catania, Italy}
\author{D.~Carbone}
\author{M.~Cavallaro}
\author{C.~Agodi}
\affiliation{Istituto Nazionale di Fisica Nucleare, Laboratori Nazionali del Sud, I-95125 Catania, Italy}
\author{J.~Lubian}
\affiliation{Instituto de F\'isica, Universidade Federal Fluminense, 24210-340, Niter\'oi, Rio de Janeiro, Brazil}
\author{J.R.B.~Oliveira}
\affiliation{Instituto de F\'isica, Universidade de S\~ao Paulo, S\~ao Paulo, Brazil}

\date{\today}

\begin{abstract} 

The nuclear scattering at energies well above the Coulomb barrier allows for a fairly sensitive examination of the parameters of the optical potential when the relevant couplings are included into the calculations. In this work we present experimental angular distribution data for the elastic and inelastic scatterings of $^{16}$O impinging on $^{27}$Al and $^{28}$Si target nuclei at \textit{E}$_{lab}=240$ MeV. The experimental data were measured at $7\degree < \theta_{c.m.} < 16\degree$ with good angular resolution. Experimental data are compared with coupled channel calculations with the inclusion of couplings to excited states in the target and projectile. We show that the shape of angular distributions are sensitive to the mass diffuseness parameter and the best agreement is achieved for $a = 0.62$ fm.  

\end{abstract}


\maketitle

\section{Introduction}
\label{Intro}

High precision measurements of the elastic scattering may provide reliable information about the optical potential (OP) in the interaction of heavy nuclei. In general, the angular distribution of the elastic cross sections exhibits a characteristic fall-off at backward angles which is interpreted as Fresnel or Fraunhofer diffraction patterns in the wave-optical description of the scattering \cite{FrL96}. Besides, the scattering of $\alpha$-cluster nuclei can exhibit more interesting features. For instance, the elastic scattering of $^{12}$C + $^{16}$O at 80 MeV \cite{GBO73} has large cross sections at backwards angles, known as anomalous large angle scattering and later identified as an interference from the indistinguishable $\alpha$-transfer reaction \cite{Dev73}. At higher beam energies, the elastic scattering exhibits local minima in the angular distributions associated to Airy minima and the interpretation of nuclear rainbow-like structure in nuclear collisions \cite{OGT00, KOB07}.

The appearance of nuclear rainbow in the elastic scattering requires a low depth of the imaginary term in the OP (transparent systems). The details of the nuclear rainbow have been exploited to remove ambiguities in the OP parameters \cite{KOB07}. The elastic scattering of $^{12}$C + $^{16}$O at 330 MeV \cite{OhH14} has been revisited in the context of an extended double-folding model. The inclusion of $3_{1}^{-}$ and $2_{1}^{+}$ states in $^{16}$O in the calculations were crucial to fully describe the angular distribution and highlights the appearance of a secondary nuclear rainbow due to coupling with excited states. This is referred to as a nuclear rainbow \cite{MHO15} generated by the dynamics of the couplings to excited states. A similar analysis has been performed for the scattering of $^{13}$C + $^{16}$O system that also points to the presence of dynamical rainbow in the elastic scattering \cite{OHO15}.

Evidences of the dynamical rainbow-like structure in the elastic scattering of $^{16}$O + $^{27}$Al system at 100 MeV and 280 MeV have been reported in Refs.~\cite{PLO12, CNL16}, respectively. Theoretical calculations of the elastic cross sections have been performed using a double-folding based OP, with the imaginary factor that effectively takes into account the loss of flux to dissipative processes and inclusion of $^{27}$Al low-lying states in the coupling matrix. This theoretical recipe gave a reasonable description of both the elastic and inelastic scatterings at 100 MeV, but it was not sufficient to explain the scattering at higher beam energies, where it was required an attenuation of the real part of the OP \cite{CNL16}. Couplings to peripheral reactions like p-transfer and $\alpha$-transfer have not improved the overall agreement between experimental data and theoretical curves.

In a recent work, the elastic and inelastic scatterings of $^{16}$O + $^{60}$Ni have been measured at 260 MeV \cite{ZCL18}. Theoretical calculations considered couplings to excited states in $^{16}$O and $^{60}$Ni by means of an imaginary potential deformation according to Refs.~\cite{Sat70,Sat83}. These have shown to be crucial for a good description of the elastic and inelastic scattering in both $^{16}$O + $^{60}$Ni at 260 MeV and $^{16}$O + $^{27}$Al at 280 MeV. The $3^-$ excited state in $^{16}$O nucleus seems to play an important role in the elastic scattering on heavy targets even at bombarding energies well above the Coulomb barrier.

In this work we perform a further investigation of the role of excited states of $^{16}$O in the scattering by $^{27}$Al and the isotone $^{28}$Si at 240 MeV. Angular distributions of elastic and inelastic cross sections have been measured at very forward angles ($4\degree < \theta_{lab} < 12\degree$) with high angular resolution. Theoretical calculations take into account couplings to excited states in both projectile and target nuclei also considering the deformation of the imaginary potential, following the method indicated in Ref.~\cite{ZCL18}.

This paper is organized as follows: the experimental details and the theoretical analysis are discussed in sections~\ref{exp} and~\ref{sec:Theoretical-Calculations}, respectively. The conclusions are given in section~\ref{conc}.

\section{\label{exp}Experimental details}

The measurements were performed at the \textit{Istituto Nazionale di Fisica Nucleare - Laboratori Nazionali del Sud}, Catania, Italy. The 240 MeV $^{16}$O$^{6+}$ beam was delivered by the superconducting cyclotron. A  $^{27}$Al (89 $\mu$g/cm$^{2}$ thickness) and $^{28}$Si (148 $\mu$g/cm$^{2}$ thickness) self-supporting foils, produced by evaporation, were used as targets. The target thickness was estimated by scanning the thin film with a collimated $\alpha$-source and measuring the residual energy of the emerging $\alpha$ particles.

\begin{figure}[tb!]
\centering
\graphicspath{{figuras/}}
\includegraphics[width=0.45\textwidth]{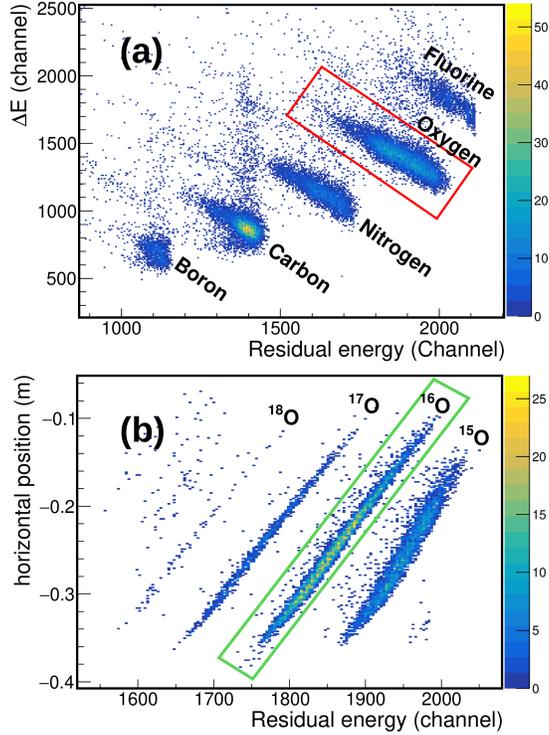}
\caption{(Color online) Typical plots for particle identification performed at the FPD of the MAGNEX spectrometer for the $^{16}$O + $^{27}$Al system. Atomic number of ejectiles are selected in a E-$\Delta$E spectra (see Fig.~\ref{spectrum}a) and projected into an horizontal position \textit{versus} residual energy plot to select the $^{16}$O isotope (Fig.~\ref{spectrum}b). All oxygen isotopes in Fig.~\ref{spectrum}b are in the $8^{+}$ charge state.} 
\label{spectrum}
\end{figure}

After traversing the target, $^{16}$O$^{8+}$ ejectiles from the reaction were momentum analyzed by the MAGNEX spectrometer \cite{LCC07,LCC08,CCC07,CaC16} set in the full acceptance mode ($\Omega \sim 50$ msr). Parameters of the final trajectory (i.e. vertical and horizontal positions and incident angles) were measured by the focal plane detector (FPD) that also allows for particle identification \cite{CCC10}. Typical spectra for particle identification in the FPD are shown in Fig.~\ref{spectrum}, obtained in the measurements with the $^{27}$Al target nucleus. Similar plots are observed in the measurements with $^{28}$Si target. The oxygen particles were selected in a standard E-$\Delta$E plot (red graphical selection in Fig.~\ref{spectrum}a). Oxygen isotopes, within the graphical selection, are identified in a position-to-energy correlation plot (Fig.~\ref{spectrum}b), which allows for a clear distinction between the $^{16}$O$^{8+}$ (elastic and inelastic scatterings), the $^{15}$O$^{8+}$ (1n stripping reaction) and the $^{17}$O$^{8+}$ (1n pick-up) and some events in the $^{18}$O$^{8+}$ (2n pick-up).

\begin{figure}[tb!]
\centering
\graphicspath{{figuras/}}
\includegraphics[width=0.45\textwidth]{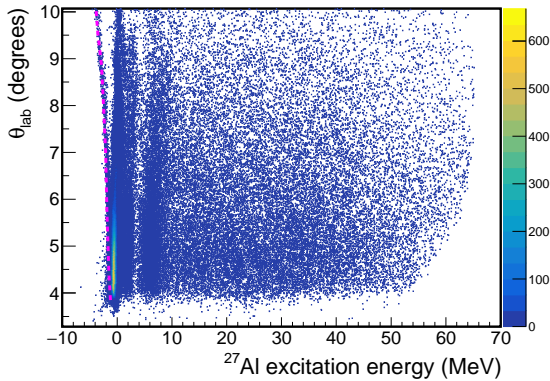}
\caption{(Color online) The $\theta_{\textnormal{lab}}$ \textit{versus} $^{27}$Al excitation energy 2d-histogram for the $^{16}$O + $^{27}$Al system. The dashed purple line shows the kinematics for $^{16}$O particles scattered by a contaminant present in the target.} 
\label{theta-ecc}
\end{figure}

The scattering data were collected at one angular setting, with the spectrometer optical axis centered at $\theta_{\textnormal{lab}} = 8\degree$. Due to the large angular acceptance of the spectrometer, this angular setting allows to cover angular range of $4\degree < \theta_{\textnormal{lab}} < 12\degree$. Trajectory reconstruction of $^{16}$O ejectiles was performed by solving the equation of motion for each particle through the magnetic fields of dipole and quadrupole, obtaining scattering parameters relative to the target position, such as the momentum vector (scattering angle $\theta_{\textnormal{lab}}$ and excitation energy relative to the residual nucleus). The overall angular resolution achieved in this measurements is better than $0.6\degree$. Further details of this procedure are found in Refs. \cite{LCC07,LCC08,CAB14,Car15}. The $\theta_{\textnormal{lab}}$ \textit{versus} $^{27}$Al excitation energy 2d-histogram for the $^{16}$O + $^{27}$Al system is shown in Fig.~\ref{theta-ecc}, in which the elastic channel corresponds to the vertical locus at 0 MeV. Inelastic scatterings, leading to excitation of target and/or projectile nuclei, produce similar loci at positive $^{27}$Al excitation energies. We observe that low-lying states are relatively well populated up to $\sim 8$ MeV, followed by an almost continuous population of states. Moreover, it must be noted a contribution due to the scattering of the beam on heavy contaminants of the target that interferes with elastic yields at $\theta_{lab}<5.5\degree$. This is indicated by a dashed purple line in Fig.~\ref{theta-ecc}. Kinematical parameters (kinematic energy and scattering angle) of this heavy contaminant are consistent with Fe, possibly introduced during production of the thin films. A similar behavior is also observed in the trajectory reconstruction for the measurements with $^{28}$Si target.

\begin{figure}[tb!]
\centering
\graphicspath{{figuras/}}
\includegraphics[width=0.45\textwidth]{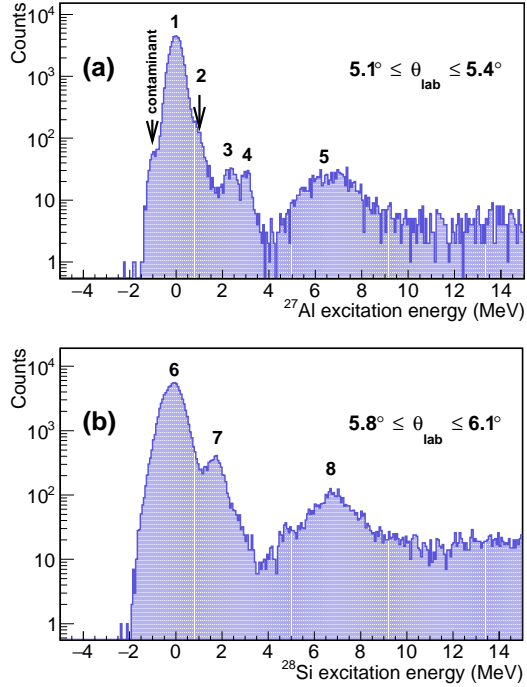}
\caption{(Color online) Typical excitation energy spectra for elastic and inelastic scattering of $^{16}$O on (a) the $^{27}$Al and (b) $^{28}$Si target nuclei. In (a), the tiny peak 2 relative to the 1/2$^{+}_{1}$ (0.81 MeV) and the 3/2$^{+}_{1}$ (1.01 MeV) states are in the tail of the elastic peak (number 1); peak 3 corresponds to the 7/2$^{+}_{1}$ (2.21 MeV); peak 4 to the sum of 5/2$^{+}_{2}$ (2.73 MeV), 3/2$^{+}_{2}$ (2.98 MeV) and 9/2$^{+}_{1}$ (3.00 MeV). In (b), the elastic peak is well resolved from $2^{+}_{1}$ state (1.78 MeV), respectively labelled as 6 and 7. In both spectra is observed a broad peak (labelled as 5 and 8) that corresponds to excitation of $^{16}$O projectile along with states in the target nuclei.} 
\label{ecc-spectra}
\end{figure}

The excitation energy spectra for $^{27}$Al and $^{28}$Si are shown in Fig.~\ref{ecc-spectra}. The main low-lying states in each target nuclei are numbered. The elastic scattering corresponds to the peaks 1 and 6 for $^{27}$Al and $^{28}$Si, respectively. The overall energy resolution is about 0.6 MeV, estimated from the full width half maximum of gaussian curves fitted to the elastic peak. In the $^{27}$Al spectrum (Fig.~\ref{ecc-spectra}a), the 1/2$^{+}_{1}$ (0.81 MeV) and the 3/2$^{+}_{1}$ (1.01 MeV) states (peak 2) are superimposed to the tail of the elastic peak. The small bump on the left of the elastic peak is due to $^{16}$O ejectiles scattered by the heavy contaminant present in the target. The other identified peaks are associated with the 7/2$^{+}_{1}$ (peak 3) and to the sum of 5/2$^{+}_{2}$, 3/2$^{+}_{2}$ and 9/2$^{+}_{1}$ (peak 4) states. These low-lying states in $^{27}$Al are the quintuplet generated by the 1d$_{5/2}$ proton hole coupled to the $2^{+}$ rotational state in $^{28}$Si core. In the $^{28}$Si spectrum (Fig.~\ref{ecc-spectra}b), the $2^{+}$ state (peak 7) is well resolved from the elastic (peak 6). There is also a small contribution from a heavier contaminant in the target underneath the elastic peak as in the $^{27}$Al spectrum. The $4^+$ ($4.62$ MeV) and $0^+$ ($4.98$ MeV) states are suppressed compared to the $2^+$. Moreover, in both spectra a bump is observed at 6.0 - 6.5 MeV (peaks 5 and 8) which is interpreted as excitation of $^{16}$O projectile followed by $\gamma$-emission in-flight superimposed to high-lying states in the target nuclei. 

\begin{figure}[tb!]
\centering
\graphicspath{{figuras/}}
\includegraphics[width=0.45\textwidth]{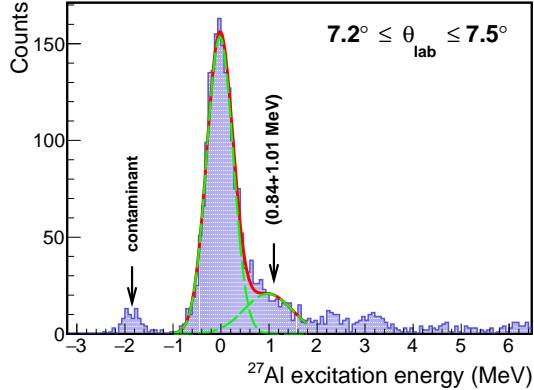}
\caption{(Color online) Typical fits to the experimental spectrum in $^{27}$Al. The 1/2$^{+}_{1}$ (0.81 MeV) and the 3/2$^{+}_{1}$ (1.01 MeV) states have been fitted to a single gaussian shape (dashed green curve). The solid red line corresponds to the sum of the two gaussian curves.} 
\label{fit-spectrum}
\end{figure}

For each angular step, the yields in the elastic and inelastic peaks have been extracted from gaussian fits to the excitation energy spectra. An example of such fitting procedure is shown in Fig.~\ref{fit-spectrum} for the $^{27}$Al case at $7.2\degree < \theta_{lab} < 7.5\degree$. The 0.84 MeV and 1.01 MeV states were fitted to a single gaussian curve due to energy resolution attained in the experiment. The width of the two gaussian curves (dashed green) were restricted to the experimental conditions. No background was considered in these fits. 

The peak associated with the contaminant in the target has been fitted independently for the angular steps in which it was clearly separated from the elastic peak. The contaminant yields in this way have been extrapolated to low angles assuming a Rutherford-like elastic scattering curve. These extrapolated yields have been subtracted from the elastic peak of the angular bins in which they are not fully resolved.

The error bars in the experimental cross sections correspond to uncertainty in the solid angle determination and counting statistics. A systematic uncertainty in the cross section of 10$\%$, coming from uncertainties in the target thickness and beam integration by the Faraday cup, is common to all the angular distribution points and is not included in the error bars.

\section{Calculations and Discussion\label{sec:Theoretical-Calculations}}

The elastic and inelastic scattering of $^{16}$O on $^{27}$Al and $^{28}$Si were measured at bombarding energy $\sim 6.8$ times higher than the Coulomb barrier. At this energy many reaction channels are open, such as nucleonic transfer reactions which produce many ejectile species (see Fig.~\ref{spectrum}). The number of direct reaction channels to be incorporated in a full coupled reaction calculation can be computationally prohibitive. Transfer reactions of few nucleons take place at the peripheral region of the nuclei and may be theoretically described by coupled reaction channels (CRC) calculations. Inelastic channels are well described by coupled channel (CC) calculations and account for the collective excitations of the nuclei. Here we focus mainly on static effects (due to the deformation of target nuclei) and dynamical effects emerging from the coupling of inelastic channels. The incorporation of possible resonant effects, via the method discussed in \cite{ZCL18}, will also be applied. The main transfer reactions will be studied in a forthcoming work.

The $^{27}$Al and $^{28}$Si nuclei are isotones that differ from each other just by one proton and, to some extent, their nuclear structures share some similarity. The first low-lying states in $^{27}$Al nucleus, given by $1/2^+$ (0.84 MeV), $3/2^+$ (1.01 MeV), $7/2^+$ (2.21 MeV), $5/2^+$ (2.73 MeV) and $9/2^+$ (3.00 MeV), are interpreted in a weak-coupling scheme as a proton hole in the $1d_{5/2}$ sub-shell built on the $2^+$ rotational state of the $^{28}$Si core. On the other hand, the energy levels spectrum of $^{28}$Si do not follow a simple rotational model \cite{WiM73}. Here we are treating these nuclear systems in the same theoretical grounds to investigate the effects of couplings, potential parameters and static deformations on the elastic and inelastic cross section angular distribution.

All calculations have been performed using the FRESCO code \cite{Tho88} and convergence was achieved considering a matching radius of $20$ fm and $300$ partial waves. The theoretical curves shown here take into account the experimental angular resolution. For the nuclear optical potential we adopted $U(r) = [N_r + i \cdot N_i] \times V_{SPP}(r)$, where $V_{SPP}(r)$ is the S\~ao Paulo potential (SPP) \cite{CCP78,CPH97,CCG02}. It was verified that this SPP-based optical potential, with normalization factors of $N_r = 1.0$ and $N_i = 0.78$, provides a reasonable description of the elastic scattering for a large set of systems \citep{ACH03}, and it is an useful starting-point for our calculations. As the nuclei involved in the present reactions present large deformation parameters, we set the calculations in order to guarantee the volume conservation up to second-order correction.

\subsection{Coupling Effects \label{sec:acoplamentos}}

\begin{figure}[H]
\centering
\includegraphics[width=0.45\textwidth]{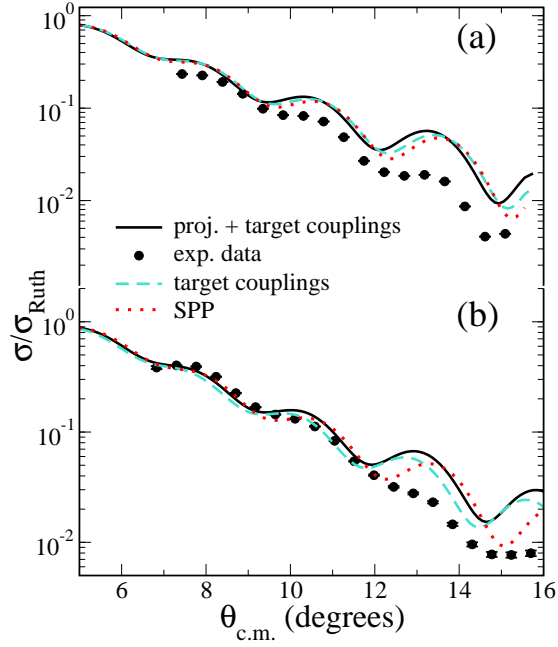} 
\caption{\textsl{\small{}{}(Color online) - Elastic angular distributions of  $^{16}$O + $^{27}$Al (a) and $^{16}$O + $^{28}$Si (b) systems at $240$ MeV: The dotted red line corresponds to the optical model calculation with the nuclear S\~ao Paulo potential with $1.0$ and $0.78$ normalization factors for the real and imaginary parts, respectively. CC calculations including inelastic states of the target only (dashed cyan line) and projectile + target excitations (black line) are also shown.}}
\label{couplingFig} 
\end{figure}

To assess the effect of couplings in the elastic scattering we first performed an optical model calculation using the SPP-based optical potential with  $N_r = 1.0$ and $N_i = 0.78$, as previously mentioned. The comparison of such optical model with experimental data in $^{27}$Al and $^{28}$Si is shown in Fig.~\ref{couplingFig} (dotted red line). For both systems, the same result is observed: the angular distributions are not well described by this calculation, specially at scattering angles $\theta_{c.m.}>10^{\circ}$, in which the experimental minima positions and, mainly, the amplitude of oscillations are not well reproduced.

In the next step, we have included inelastic channels associated to target excitations. For the $^{27}$Al target, the low-lying states $1/2^+$ (0.84 MeV), $3/2^+$ (1.01 MeV), $7/2^+$ (2.21 MeV), $5/2^+$ (2.73 MeV) and $9/2^+$ (3.00 MeV) were considered. Some of them are clearly observed in the excitation energy spectrum (see Fig.~\ref{ecc-spectra}a). For the $^{28}$Si target, we  included the $2^{+}$ state ($1.78$ MeV), observed in the excitation energy spectrum (see Fig.~\ref{ecc-spectra}b). The intrinsic matrix elements for these transitions were calculated from the experimental reduced transition probabilities $B(E2)$ and are listed in Table \ref{tab:be2}. We have also included the $4^+$ and $0^+$ in $^{28}$Si. The calculations assuming the couplings to excited states of the target are indicated as dashed cyan lines in Fig.~\ref{couplingFig}. The parameters of the OP were kept as before. The effects of coupling with excited states in the target is to introduces a small displacement on the phase at $\theta_{c.m.} > 10^{\circ}$. However, the couplings to the inelastic channels of target are still not sufficient to properly describe the data.

\begin{table}[t]
\caption{\textsl{\small{}{}Experimental reduced transition probabilities $B(E2)$ for the excited states in $^{27}$Al and $^{28}$Si adopted for the couplings to inelastic channels. Values for transitions in $^{27}$Al are from Ref.~\cite{Sha11}. For the $2^+$ in $^{28}$Si, $B(E2) \uparrow$ are from Ref.~\cite{RNT01} and for the $4^+$ and $0^+$ $B(E2) \downarrow$ are from Ref.~\citep{Sha13}  }}
\resizebox{0.95\textwidth}{!}{\begin{minipage}{\textwidth}
\begin{tabular}{cccccc}
\hline
\hline 
\multicolumn{3}{c}{\textbf{states in $^{27}$Al}} & \multicolumn{3}{c}{\textbf{states in $^{28}$Si}}  \\
\textbf{initial} & \textbf{final} & \textbf{$B(E2)$ $(e^2 b^2)$} & \textbf{initial} & \textbf{final} & \textbf{$B(E2)$ $(e^2 b^2)$} \\
\hline
$5/2^+$ & $1/2^+$  & $0.004$ & $0^+$  & $2^+$  & $0.033$   \\  
$5/2^+$ & $3/2^+$  & $0.019$ & $2^+$  & $4^+$  & $0.008$   \\  
$5/2^+$ & $7/2^+$  & $0.004$ & $2^+$  & $0^+$  & $0.005$   \\  
$5/2^+$ & $5/2^+$  & $0.007$ &        &       &     \\  
$5/2^+$ & $9/2^+$  & $0.004$ &        &       &     \\  
\hline
\end{tabular}
\label{tab:be2}
\end{minipage}}
\end{table}

Recently, in Ref.~\cite{ZCL18} it has been pointed out that the excitation of $^{16}$O projectile may play an important role in the scattering of nuclei at energies well above the Coulomb barrier. In the excitation energy spectra (Fig.~\ref{ecc-spectra}), a peak roughly at $6$ MeV appears in both systems that is possibly related to the excitation of the projectile, namely, the $3^-$ state of $^{16}$O with $E^{\ast}=6.1$ MeV. This channel has been added to the previous coupling scheme (considering the $0^+ \rightarrow 3^-$ transition) using the $B(E3) \uparrow$ value of $0.0015$ $e^2b^3$ reported in Ref.\citep{KiS02}. The results of such coupling may be seen in Fig.~\ref{couplingFig} as the black line. Once again, the effect of this channel is to introduce a small shift in the minima positions at the backward region. Once again, the inclusion of inelastic channels is not sufficient to damp the oscillations seen in the calculations. The inclusion of transfer reaction channels (p-transfer and $\alpha$-transfer) does not appreciable change the calculations shown so far and we are not showing here.

\subsection{Potential Effects  \label{sec:potenciais}}              

\begin{figure}[H]
\centering
\includegraphics[width=0.45\textwidth]{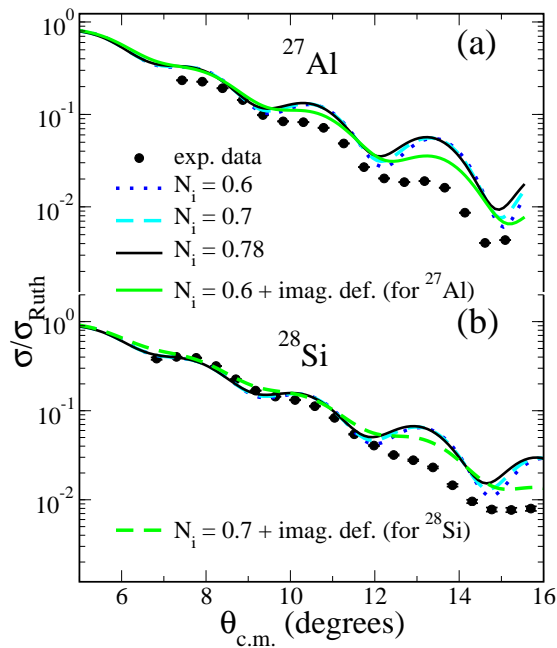} 
\caption{\textsl{\small{}{}(Color online) - Elastic angular distributions of $^{16}$O + $^{27}$Al (a) and $^{16}$O + $^{28}$Si (b) systems at $240$ MeV with full CC calculation using $N_i=0.78$ (black lines), $N_i=0.7$ (dashed light blue line) and $N_i=0.6$ (dotted blue line). The green lines correspond to calculations changing the mass diffuseness.}}
\label{fig:potential} 
\end{figure}

Considering all couplings to excited states of the target and projectile nuclei above mentioned, we examined the effects of the imaginary normalization factor ($N_i$) in the angular distribution for the elastic scattering. The usual value, according to a systematic analysis performed in Ref.~\cite{ACH03}, is $N_i = 0.78$ and this calculation is shown in dotted red lines in Fig.~\ref{couplingFig}.  The explicit inclusion of inelastic channels in the coupling scheme may require an attenuation of the absorption in the OP. We performed same calculations varying the values for $N_i$. In Fig.~\ref{fig:potential} we only show the curves for CC calculation with $N_i=0.6$ and $0.7$. For reference we also show the black line from Fig.~\ref{couplingFig}. The imaginary factor does not result in a perceptibly improvement in the agreement between calculation and data. We just observe a small decrease of the amplitude of oscillation at $\theta > 10\degree$ and no shifts in the minima angles.

At high bombarding energies, highly collective states of the target nucleus can be populated. The inclusion of couplings to these states is still an open topic for the theory of direct reactions. In Ref.~\cite{ZCL18} it was discussed a possibility to effectively take these couplings into account in a CC calculation using a deformed complex coupling potential, inspired by the Bohr-Mottelson unified model \cite{ABH56,HiW53}. For the deformation of the imaginary part of the OP, we carried out a multipole expansion up to the octupole term, just as it is made to the deformation of the real part of the potential. The calculation with deformation is shown in green lines of Fig.~\ref{fig:potential}. The best correspondence of data to theoretical calculations is achieved for imaginary normalization of $N_i=0.6$ (for the $^{27}$Al) and $N_i=0.7$ ($^{28}$Si). The deformation of the imaginary term of the OP reduces the amplitude of oscillation in the calculations, barely modifying the position of minima. 

As mentioned before, the transfer reaction channels (not shown here) do not improve the agreement between calculations and data. Dynamical and potential effects seems not sufficient to describe our experimental data so we proceed to study the static effects.

\subsection{Static Effects  \label{sec:estaticos}}       

It is well known that both target nuclei are deformed, as indicated by their high electric quadrupole moment: $Q=+0.14(1) b$ for $^{27}$Al and $Q=+0.16(1) b$ for $^{28}$Si, according to \cite{Sto05}. One way to include such intrinsic deformation is to adjust the nuclear mass diffuseness parameter of the OP. In the systematics for the SPP, this value was set to $a=0.56$ fm. It is important to mention that this systematic values for the nuclear mass distribution is based on 2 parameters for the Fermi distribution that is considered to be spherical. A way to effectively account for the deformation of the matter density in the ground state might be achieved by changing the radius or the diffuseness of the matter distribution. In Refs.~\cite{COS11,CZS18} the deformation has been treated by changing the mass diffuseness. In these works, the quasi-elastic barrier distribution has been studied and it was shown that on $^{18}$O$+^{60}$Ni, $^{18}$O$+^{63}$Cu system, the centroid of the barrier was very sensitive to the increment of the matter density of the $^{18}$O, considered as $^{16}$O core plus 2 neutrons. For this reason, these 2 extra neutrons produce a matter density that is more diffuse than the usual nuclei, as is the case also in halo nuclei, in a smaller scale. The same situation was also observed in the reaction involving $^{17}$O projectile due to the same reasons.

\begin{figure}[H]
\centering
\includegraphics[width=0.45\textwidth]{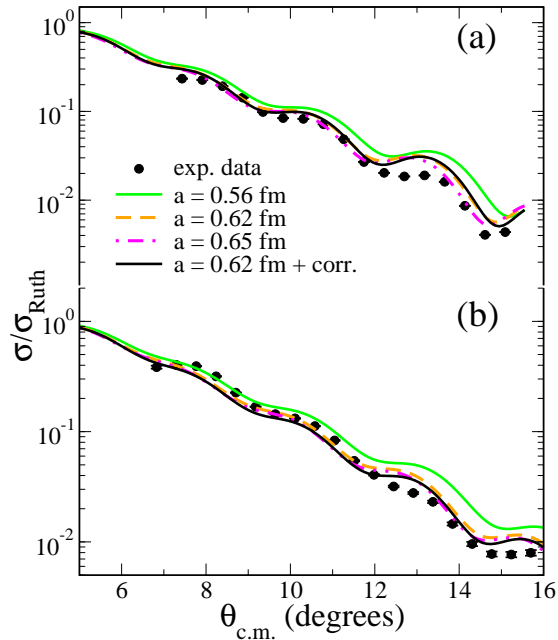} 
\caption{\textsl{\small{}{}(Color online) - Elastic angular distributions of $^{16}$O + $^{27}$Al (a) and $^{16}$O + $^{28}$Si (b) systems at $240$ MeV: The different lines are results of different nuclear mass diffuseness parameters adopted in the calculations. The black lines show the effect of incorporating the finite diffuseness value of nuclear density.}}
\label{fig:staticA} 
\end{figure}

A similar approach has been used in the present work, studying effect of the nuclear mass diffuseness parameter to the angular distributions. The results are shown in Fig.~\ref{fig:staticA}. 
The calculations performed in this section used the couplings discussed in section \ref{sec:acoplamentos}. The elastic angular distribution is quite sensitive in the backward angular region to the variation of mass diffuseness. The calculations definitely show that, for a proper description of the distribution, the static deformation of nuclei must be incorporated on calculations (via the diffuseness parameter in the optical potential on this case). The comparison of data with calculations showed that such diffuseness should be between $a=0.62$ and $0.65$ fm. We also included second-order corrections for the deformation based on the generalized rotation-vibration model as described in Ref.~\cite{ChC10}, that dealt with the effect of the finite diffuseness value of the nuclear density, not observed in pure vibrational or rotational models. The effect of such corrections, applied to the calculation with $a = 0.62$ fm, is represented by the black curve in Fig.~\ref{fig:staticA}.


\begin{figure}[H]
\centering
\includegraphics[width=0.45\textwidth]{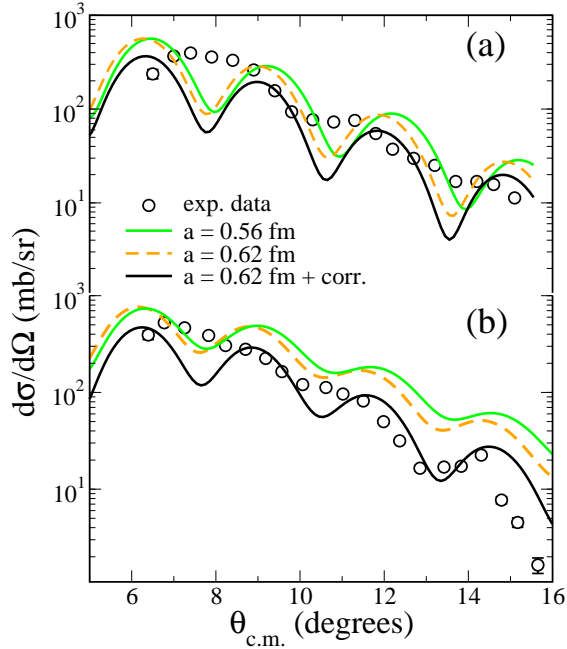} 
\caption{\textsl{\small{}{}(Color online) - Inelastic angular distributions of $^{16}$O + $^{27}$Al (a) and $^{16}$O + $^{28}$Si (b) systems at $240$ MeV: The  black line shows the effect of incorporating the finite diffuseness value of nuclear density, while the dashed orange line does not incorporate such effect on calculations. For the $^{27}$Al, we include the sum of all low-lying states labeled as peaks 2-4, in Fig.~\ref{ecc-spectra}. }}
\label{fig:staticB} 
\end{figure}      

\begin{figure}[H]
\centering
\includegraphics[width=0.45\textwidth]{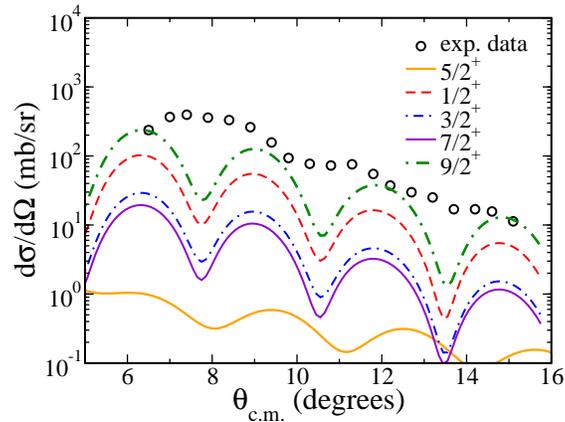} 
\caption{\textsl{\small{}{}(Color online) - Individual inelastic angular distributions of $^{16}$O + $^{27}$Al for the low-lying states in $^{27}$Al.}}
\label{fig:sumInelastic} 
\end{figure}      

Fig.~\ref{fig:staticB} shows the inelastic angular distribution of both systems. For the $^{27}$Al, the experimental cross sections correspond to the sum of the 5 low-lying states, and for the $^{28}$Si it is just the first $2^+$ excited state. The minima observed in the angular distributions are more pronounced in the $^{27}$Al than in $^{28}$Si because the individual inelastic cross sections for $1/2^+$, $3/2^+$, $7/2^+$ and $9/2^+$ excited states (in $^{27}$Al) oscillates in phase, as expected for the coupling of a 1d$5/2$ proton hole to a $^{28}$Si core (see Fig.~\ref{fig:sumInelastic}). In addition, the coulomb interaction is stronger for $^{28}$Si than for $^{27}$Al and causes the attenuation in the minima for $^{28}$Si compared to the $^{27}$Al. The calculations with the usual adopted mass diffuseness parameter ($a = 0.56$ fm) is represented by the green curves. We also show the calculation for mass diffuseness parameter set to $a=0.62$ fm (dashed orange curves). In both systems these calculations overestimate the experimental data. However, when we include the finite diffuseness correction a slightly better agreement is observed (black line in Fig.~\ref{fig:staticB}). The oscillations of the $^{27}$Al inelastic states are not so well reproduced as those of the silicon case, probably because the separation of these states and the elastic channel on the experiment was not so good and some contamination may occur. One may also observe on Fig.~\ref{fig:staticA} that such correction results in a small change on the elastic scattering. Even there, data and theoretical lines are compatible.     

\section{\label{conc}Conclusions}
  
To summarize, in this work we presented new experimental data for the cross section of elastic and inelastic scatterings of the $^{16}$O by the isotones $^{27}$Al and $^{28}$Si nuclei at $E_{lab} = 240$ MeV. The high accuracy and precision of the data allowed us to analyze in detail the OP models adopted for these systems. To obtain a better agreement with experimental data we need to apply a twofold procedure: i) consider the deformation of the imaginary term of the OP and ii) operating fine tuning of the mass diffuseness parameter. In the present case we concluded that the angular distributions of the cross sections for the elastic and inelastic scattering are well described using $a=0.62$ fm, which is slightly higher than the usual values in the systematics of the SPP ($a=0.56$ fm).

\section*{Acknowledgment}
This project has received funding from the European Research Council (ERC) under the European Union’s Horizon 2020 research and innovation programme (grant agreement No 714625). The Brazilian authors acknowledgment partial financial support from CNPq (Proc. No. $464898$/$2014-5$), FAPERJ, FAPESP, CAPES and from INCT-FNA (Instituto Nacional de Ci\^ {e}ncia e Tecnologia- F\' isica Nuclear e Aplica\c {c}\~ {o}es).

\pagebreak


%


\end{document}